\documentstyle[aps,epsfig]{revtex}
\newcommand{\lsim}{\mathrel{\rlap{\lower4pt\hbox{\hskip0pt$\sim$}}
\raise1pt\hbox{$<$}}}
\newcommand{\gsim}{\mathrel{\rlap{\lower4pt\hbox{\hskip0pt$\sim$}}
\raise1pt\hbox{$>$}}}
\newcommand{\sfrac}[2]{\mbox{\footnotesize $\frac{#1}{#2}$}}
\begin{document}
\twocolumn

\title{Valence-quark distributions in the pion}
%
\author{M.B.\ Hecht, C.D.\ Roberts and S.M.\ Schmidt\vspace*{0.4em}}
\address{Physics Division, Argonne National Laboratory, Argonne IL
60439-4843\\[0.6\baselineskip]
\parbox{140mm}{\rm \hspace*{1.0em} 
We calculate the pion's valence-quark momentum-fraction probability
distribution using a Dyson-Schwinger equation model.  Valence-quarks with an
active mass of $0.30\,$GeV carry $71$\% of the pion's momentum at a resolving
scale $q_0=0.54\,{\rm GeV}=1/(0.37\,{\rm fm})$.  The shape of the calculated
distribution is characteristic of a strongly bound system and, evolved from
$q_0$ to $q=2\,$GeV, it yields first, second and third moments in agreement
with lattice and phenomenological estimates, and valence-quarks carrying
$49$\% of the pion's momentum.  However, pointwise there is a discrepancy
between our calculated distribution and that hitherto inferred from
parametrisations of extant pion-nucleon Drell-Yan data.\\[0.4\baselineskip]
Pacs Numbers: 13.60.Hb, 14.40.Aq, 12.38.Lg, 24.85.+p
\vspace*{8ex}
%
}}
%
%
%
\maketitle


\section{Introduction}
The cross section for deep inelastic lepton-hadron scattering can be
interpreted in terms of the momentum-fraction probability distributions of
quarks and gluons (partons) in the hadronic target, and since the pion is a
two-body bound state with only $u$- and $d$-valence-quarks it is, in some
respects, the simplest hadron and therefore represents the least complicated,
nontrivial system for which these distribution functions can be calculated.
In a theorist's ideal world, stable pion targets would then facilitate a
comparison between these calculations and the parton distribution functions
measured in deep inelastic scattering.  However, pion targets are not
abundant, and therefore these functions have primarily been inferred from
Drell-Yan\cite{DYexp,DYexp2} and direct photon production\cite{Dgammaexp} in
pion-nucleon and pion-nucleus collisions: an approach that can only be
successful if the nucleon's parton distributions are well known\cite{sutton}.
More recently, semi-inclusive e$\,$p $\to$ e$\,NX$ reactions, $N=p,n$, have
also been employed\cite{HERA}, a method which assumes that small-virtuality
pions dominate leading nucleon production.  These experiments are challenging
but high-statistics data exist.

The theoretical situation is arguably poorer: the systematic tool provided by
perturbative QCD does not admit the calculation of the distribution functions
but only of their evolution from one large spacelike-$q^2$ to another; and
numerical simulations of lattice-QCD are currently restricted to the quenched
approximation and only yield moments of the distributions, not the
distributions themselves\cite{lattice}.  Furthermore, the fact that the pion
is both a bound state and the Goldstone mode associated with dynamical chiral
symmetry breaking complicates the calculation of the pion's distribution
functions and places additional constraints on any framework applied to the
task.  Considerations of chiral symmetry have led
some\cite{toki,arriola,bentz} to adopt the Nambu--Jona-Lasinio model as a
basis for their calculations but a number of the difficulties with this
approach, among them a marked sensitivity to the regularisation procedure in
this non-renormalisable model, are emphasised and discussed in
Refs.\cite{arriola,bentz}.  Constituent quark models have also been
employed\cite{cotanchmiller}, with difficulties encountered in such studies
considered in Ref.\cite{shakin}; and so has an instanton-liquid
model\cite{dorokhov}.

The Dyson-Schwinger equations (DSEs)\cite{cdragw} provide an approach
well-suited to the calculation of pion observables.  Since a chiral symmetry
preserving truncation scheme exists\cite{truncscheme}, they easily capture
the dichotomous bound-state/Goldstone-mode character of the
pion\cite{mrt98,mr97}.  Furthermore, because perturbation theory is recovered
in the weak coupling limit, they combine; e.g., a description of low-energy
$\pi$-$\pi$ scattering\cite{pipi} with a calculation of the electromagnetic
pion form factor, $F_\pi(q^2)$, that yields\cite{mrpion}: the
$1/q^2$-behaviour expected from perturbative analyses at large
spacelike-$q^2$ and a calculated evolution to the $\rho$-meson pole in the
timelike region\cite{mtpion}.  These and other features of contemporary
applications are described in Refs.\cite{revbasti,revreinhard}.

Herein we employ a phenomenological DSE model, used
previously\cite{jacquesnucleon,martinnucleon} in a description of a wide
range of electron-nucleon and meson-nucleon form factors, in a calculation of
the pion's valence-quark distribution.  The framework is Poincar\'e covariant
and exhibits, in a simplified form, the features of renormalisability and
asymptotic freedom characteristic of QCD.  Another important aspect of our
study is a description of the pion as a bound state of a dressed-quark and
-antiquark through its momentum-{\it dependent} Bethe-Salpeter amplitude,
with the dressed-quark propagator exhibiting that momentum-dependence
characteristic of DSE studies and recently confirmed in
simulations\cite{latticequark} of lattice-QCD.

Our article is organised thus: Sec.~II recapitulates the calculation of the
hadronic tensor for lepton-pion scattering; Sec.~III describes the DSE
elements necessary in the calculation of the distribution functions --
dressed-propagators, Bethe-Salpeter amplitudes, etc.; Sec.~IV reports and
discusses the results; and Sec.~V is an epilogue.

\begin{figure}[t]
\centering{\
\epsfig{figure=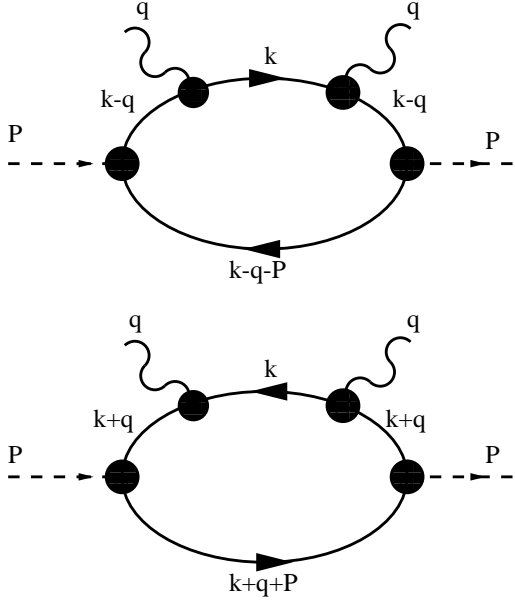,height=8.0cm}}\vspace*{0.5\baselineskip} 
\caption{\label{compton} ``Handbag'' contributions to the virtual photon-pion
forward Compton scattering amplitude, which are the only impulse
approximation diagrams that survive in the deep inelastic Bjorken limit,
Eq.~(\protect\ref{BJlim}).  $\pi$, dashed-line; $\gamma$, wavy-line; ${\cal
S}$, internal solid-line, dressed-quark propagator, Eq.\
(\protect\ref{qprop}).  The filled circles represent the pion's
Bethe-Salpeter amplitude, $\Gamma_\pi$ in Eq.\ (\protect\ref{genpibsa}), and
the dressed-quark-photon vertex, $\Gamma_\mu$ in Eq.\ (\protect\ref{bcvtx}),
depending on which external line they begin/end.}
\end{figure}
\section{Lepton-Pion Scattering}
Deep inelastic scattering from a pion target can be studied via the diagrams
in Fig.~\ref{compton}, which describe virtual photon-pion forward Compton
scattering and are\cite{landshoff} the only impulse approximation
contributions to survive in the Bjorken limit:\footnote{We use a Euclidean
metric convention, in which
$a\cdot b := a_\mu\,b_\nu\,\delta_{\mu\nu} := \sum_{i=1}^4\,a_i\,b_i\,,$
so that a spacelike vector, $Q_\mu$, has $Q^2>0$ and an on-shell pion is
described by $P^2=-m_\pi^2$.  Our Dirac matrices are Hermitian and are
defined by the algebra
$ \{\gamma_\mu,\gamma_\nu\} = 2\,\delta_{\mu\nu}$.}
\begin{equation}
\label{BJlim}
q^2\to\infty\,,\; \; P\cdot q \to -\infty \;\;\mbox{but}\;\; x:=
-\frac{q^2}{2 P\cdot q}\;\; \mbox{fixed.}
\end{equation}
The upper diagram represents the renormalised matrix element
\begin{eqnarray}
\label{Tmunu} \lefteqn{T^+_{\mu\nu}(q,P) = }\\
&& \nonumber {\rm tr}\!\int\!\frac{d^4
k}{(2\pi)^4}\,\tau_-\bar\Gamma_\pi(k_\Gamma;-P)\,S(k_t)\,
ieQ\Gamma_\nu(k_t,k) \\
&& \nonumber
\times \,S(k)\,ieQ\Gamma_\mu(k,k_t)\,S(k_t)\,
\tau_+\Gamma_\pi(k_\Gamma;P)\,S(k_s)\,,
\end{eqnarray}
where: $\Gamma_\pi(\ell;P)$ is the pion's Bethe-Salpeter amplitude and
\begin{equation}
\bar\Gamma_\pi(\ell;-P)= C^\dagger \Gamma_\pi(-\ell;-P)^{\rm T}C\,, 
\end{equation}
with $\tau_{\pm} = \case{1}{2}(\tau_1 \pm i \tau_2)$, $C=\gamma_2\gamma_4$,
the charge conjugation matrix, and $(\cdot)^{\rm T}$ denoting matrix
transpose; ${\cal S}(\ell)= {\rm diag}[S_u(\ell),S_d(\ell)]$ is the
dressed-quark propagator and we assume $S_u = S_d = S$ throughout (isospin
symmetry of the strong interaction); $\Gamma_\mu(\ell_1,\ell_2)$ is the
dressed-quark-photon vertex, with $Q= {\rm diag}(2/3,-1/3)$ the quark-charge
matrix; $k_\Gamma= k-q-P/2$, $k_t=k-q$, $k_s=k-q-P$; and the trace is over
colour, flavour and Dirac indices.  (In Eq.\ \ref{Tmunu}, since the
renormalised matrix element is finite in our DSE model, as in QCD, we have
not made explicit the usual {\it translationally invariant} regularisation
scheme\cite{mr97}.)  The matrix element represented by the lower diagram is
the crossing partner of Eq.\ (\ref{Tmunu}) and is obvious by analogy.

The hadronic tensor relevant to inclusive deep inelastic lepton-pion
scattering can be obtained from the forward Compton process via the optical
theorem:
\begin{equation}
\label{Wmunu}
W_{\mu\nu}(q;P)= \frac{1}{2\pi} {\bf\sf Im}\left[
T^+_{\mu\nu}(q;P) + T^-_{\mu\nu}(q;P)
\right]\,,
\end{equation}
and because of current conservation it may be expressed in terms of only two
invariant structure functions:
\begin{equation}
W_{\mu\nu}(q;P) = W_1(q^2,P\cdot q)\, t_{\mu\nu} - \frac{W_2(q^2,P\cdot
q)}{P\cdot q} \,P_\mu^t P_\nu^t\,,
\end{equation}
with $t_{\mu\nu} = \delta_{\mu\nu} - q_\mu q_\nu/q^2$ and $P_\mu^t = P_\mu -
q_\mu \,P\cdot q/q^2$.

To proceed it is useful to express the dressed-quark propagator as
\begin{eqnarray}
\label{SDN}
S(k) & = & \Delta(k^2)\, {\cal N}(k) \,,\\ \Delta(k^2) & = & 1/[k^2+
M^2(k^2)]\,,\\ {\cal N}(k) & = & Z(k^2)\,\left[- i \gamma\cdot k +
M(k^2)\right]\,,
\end{eqnarray}
where $Z(k^2)$ is the dressed-quark wave-function renormalisation and
$M(k^2)$ is the dressed-quark mass function.  Using Eq.\ (\ref{SDN}) and
evaluating the colour and flavour traces, Eq.\ (\ref{Tmunu}) assumes the form
\begin{eqnarray}
\nonumber
\lefteqn{T^+_{\mu\nu}(q,P) = }\\
&& e^2\,\frac{4}{9}\,N_c\int\!\frac{d^4 k}{(2\pi)^4}\,
\Delta(k^2)\,\Delta(k_s^2)\,{\cal T}^+_{\mu\nu}(k,q,P)\,,
\end{eqnarray}
where
\begin{eqnarray}
\nonumber 
\lefteqn{{\cal T}^+_{\mu\nu} = {\rm tr}_D\left[
\bar\Gamma_\pi(k_\Gamma;-P)\,S(k_{t})\,i\Gamma_\nu(k_t,k)\,
\right.}\\ 
& \times & \left. {\cal N}(k)\,
i\Gamma_\mu(k,k_t)\,S(k_{t})\,\Gamma_\pi(k_\Gamma;P)\, {\cal N}(k_{s})
\right]
\label{calT}
\end{eqnarray}
and the trace is now, of course, only over Dirac indices.  In this form it is
easy to adapt the procedure of Ref.\ \cite{landshoff} and isolate the pinch
singularities that yield the imaginary part in Eq.\ (\ref{Wmunu}).

Introducing the integration variable transformations:
\begin{equation}
\label{var1}
k = \kappa + y \,q + z\, P\,,\; \kappa\cdot q = 0 = \kappa\cdot P\,,
\end{equation}
which has Jacobian $J = P\cdot q$ in the Bjorken limit, Eq.\ (\ref{BJlim}),
and subsequently
\begin{equation}
\label{var2}
y = 1 + \frac{w}{2 \,P\cdot q} - \frac{z\,P^2}{2 \,P \cdot q}\,,
\end{equation}
one finds 
\begin{equation}
k^2 = 2\,P\cdot q\,(z - x) + {\rm O}((P\cdot q)^0)
\end{equation}
so that on the domain relevant to inclusive deep inelastic scattering
\begin{equation}
\Delta(k^2) \approx \frac{1}{2\, P\cdot q}\frac{1}{z-x} 
\to \frac{i \pi}{2\, P\cdot q}\,\delta(z-x)
\end{equation}
and the contribution of $T^+$ to $W_{\mu\nu}$ can be written
\begin{eqnarray}
\nonumber \lefteqn{W^+_{\mu\nu}(q;P)= -\frac{1}{2\pi}\,\frac{1}{4\, P\cdot
q}\,e^2\,\frac{4}{9}\,N_c\,}\\
& \times & \frac{\pi}{(2\pi)^4}\int
dw\,d^2\kappa\,\left.{\bf\sf Im}\left[\Delta(k_{s}^2)\,{\cal
T}^+_{\mu\nu}(k,q,P)\right]\right|_{z=x}\,.
\end{eqnarray}
This shows that in the Bjorken limit the struck parton carries a fraction $x$
of the pion's momentum.

The integrand can now be simplified further by using Eqs.\ (\ref{var1}) and
(\ref{var2}) to express
\begin{equation}
s:=k_{s}^2 = (x-1)\,[w - P^2 + \kappa^2/(x-1)]
\end{equation}
and hence
\begin{equation}
\Delta(s) = - \frac{1}{1-x} \, \frac{1}{w - P^2 - [\kappa^2+M^2(s)]/(1-x)}
\end{equation}
so that 
\begin{eqnarray}
\nonumber
\lefteqn{W^+_{\mu\nu}(q;P)= \frac{1}{P\cdot q}\,\frac{1}{1-x}}\\
& \times & e^2\,\frac{4}{9}\,N_c\,\frac{\pi}{(2\pi)^4}\int
d^2\kappa\,\left.\frac{1}{4}\left[{\cal
T}^+_{\mu\nu}(k,q,P)\right]\right|_{z=x}^{w=w(\kappa)}\!,
\end{eqnarray}
with $w(\kappa)= P^2+(\kappa^2+\check M^2)/(1-x)$, where the valence-quark
mass is obtained as the solution of $\check M=M(-\check M^2)$.

The momentum of the remaining dressed-quark line also simplifies: 
\begin{equation}
\mu:= k_{t}^2 = \kappa^2 + x \,w\,,
\end{equation}
which illustrates that the integrand depends only on $\kappa^2$ as an
integration variable.  Another shift of variables is now useful: $\int
d^2\kappa = \pi \int d\kappa^2$,
\begin{eqnarray}
\kappa^2 & = & (1-x)\,(\mu-\mu_{\rm min})\,,\\
\mu_{\rm min} & = & x \,[P^2 + \check M^2/(1-x)]\,,
\label{mumin}
\end{eqnarray}
and hence
\begin{eqnarray}
\nonumber
\lefteqn{W^+_{\mu\nu}(q;P)= \frac{1}{2}\,e^2\,\frac{4}{9}\,N_c\,}\\
& \times & \frac{1}{(4\pi)^2}
\int^\infty_{\mu_{\rm min}}d\mu\,\left.\frac{1}{4\,P\cdot q}\left[{\cal
T}^+_{\mu\nu}(k,q,P)\right]\right|_{z=x}^{w=w(\mu)}\!,
\label{Wnew}
\end{eqnarray}
with $w(\mu)= \mu + \mu_{\rm min} (1-x)/x$.  

It is apparent in Eq.\ (\ref{Wnew}) that the only remaining functional
dependence that is not obviously expressed solely in terms of $x$ is that
arising from the nontrivial aspects of the Dirac trace in Eq.\ (\ref{calT}).
Nevertheless, direct calculation; e.g., \cite{landshoff}, shows that in the
Bjorken limit
\begin{equation}
\left.\left[{\cal
T}^+_{\mu\nu}(k,q,P)\right]\right|_{z=x}^{w=w(\kappa)} \approx
P\cdot q\, \left( a(x)\,t_{\mu\nu} + \ldots \right)
\end{equation}
so that one has
\begin{equation}
W^+_{\mu\nu}(q;P) = F_1^+(x)\, t_{\mu\nu} - \frac{F_2^+(x)}{P\cdot q}
\,P_\mu^t P_\nu^t\,,
\end{equation}
and the Callan-Gross relation:
\begin{equation}
\label{CGreln}
F_2^+(x) = 2\, x\, F_1^+(x)\,.
\end{equation}
At this point it is obvious that 
\begin{equation}
F_{1,2}^+(x) \to 0\;{\rm as}\;x\to 1
\end{equation}
because of the contraction of the integration domain: $\mu_{\rm min}\to
\infty$ as $x\to 1$.  The analysis of $W_{\mu\nu}^-$ follows a similar
pattern and yields the same results, and $F_{1,2} = F^+_{1,2} + F^-_{1,2}$.

Equations (\ref{Wmunu}), (\ref{calT}), (\ref{Wnew}) and (\ref{CGreln})
provide a model-independent starting point for calculating the valence-quark
distribution functions using model input for the internal elements
represented in Fig.\ \ref{compton}; i.e., the dressed-quark propagator, pion
Bethe-Salpeter amplitude and dressed-quark-photon vertex.  These are
valence-quark distributions because, although sea-quarks are implicitly
contained in the dressing of the propagators and calculation of the dressed
vertices, the ``handbag'' impulse approximation diagrams in Fig.\
\ref{compton} only admit a coupling of the photon to the propagator of the
dressed-quark constituent.  The internal structure of the dressed-quark is
not resolved and therefore the calculation yields the distribution at a scale
$q_0$ characteristic of the resolution: $q_0$ is an {\it a priori}
undetermined parameter in calculations such as ours, although we anticipate
$0.3\lsim q_0\lsim 1.0\,$GeV, with the lower bound set by the Euclidean
constituent-quark mass and the upper by the onset of the perturbative domain.
A sea-quark distribution is generated via the renormalisation group
(evolution) equations when the valence distribution is evolved to that
$q^2$-scale appropriate to a given experiment.  To generate explicit
sea-quark contributions at the scale $q_0^2$ requires going beyond the
impulse (handbag) approximation; e.g., incorporating photon couplings to the
intermediate-state quark-meson-loops that can appear as a dressing of the
quark propagator: $\pi^+ = u\,\bar d \to (u \bar s s) \, \bar d = (K^+ s)\,
\bar d\to u\,\bar d = \pi^+$, with $\gamma K^+ s \to K^+ s\,\gamma$, etc.
Such intermediate states arise as vertex corrections in the quark-DSE.
%

With these observations in mind then
\begin{eqnarray}
\nonumber
F_2^+(x;q_0) & = & \case{4}{9}\,x\, u_v(x;q_0) 
= \case{4}{9}\,x\,[u(x;q_0)-\bar u(x;q_0)]\,,\\
\nonumber
F_2^-(x;q_0) & = & \case{1}{9}\,x\, \bar d_v(x;q_0)
= \case{1}{9}\,x\,[\bar d(x;q_0)- d(x;q_0)]\,,\\
&& 
\label{uvx}
\end{eqnarray}
and it is straightforward to demonstrate algebraically that
\begin{eqnarray}
\bar d_v^{\pi^+}(x;q_0) & = & u_v^{\pi^+}(x;q_0) = d_{v}^{\pi^-}(x;q_0) \,.
\end{eqnarray}
The calculations
are required to yield
\begin{equation}
\label{uvnorm}
\int_0^1\,dx\,u_v(x;q_0) = 1 = \int_0^1\,dx\,\bar d_v(x;q_0)\,;
\end{equation}
i.e., to ensure that the $\pi^+$ contains one, and only one,
$u$-valence-quark and one $\bar d$-valence-quark.  

We remark that $u^{\pi^+}(x)=\bar d^{\pi^+}(x)$ in the ${\cal G}$-parity
symmetric limit.  However, in this limit $\omega \to \pi\pi$ is forbidden and
hence the scale of ${\cal G}$-parity symmetry violation in nature is
characterised by the ratio\cite{pdg00}
$\Gamma_{\omega\to\pi\pi}/\Gamma_{\rho\to\pi\pi} = 0.1$\%.  This bound on any
difference between the pion's quark distribution functions is consistent with
the model estimate in Ref.\ \cite{thomas}.

\section{Model Elements}
\label{DSEmodel}
To complete our calculation the internal lines and irreducible vertices
appearing in Fig.\ \ref{compton} must be specified.  In general these
quantities can be obtained by solving the quark DSE and the appropriate
inhomogeneous Bethe-Salpeter equations\cite{revbasti}.  However, the study of
an extensive range of low- and high-energy light- and heavy-quark phenomena
has yielded efficacious algebraic
parametrisations\cite{jacquesnucleon,martinnucleon,mark,pichowsky,pichowsky2,mishasvy},
and we employ those herein.

The dressed-quark propagator is
\begin{eqnarray}
\label{qprop}
S(p) & = & -i\gamma\cdot p\, \sigma_V(p^2) + \sigma_S(p^2)\,,\\
& = & \label{defS}
\left[i \gamma\cdot p \, A(p^2) + B(p^2)\right]^{-1}\,,\\
\label{ssm}
\bar\sigma_S(x) & =&  2\,\bar m \,{\cal F}(2 (x+\bar m^2))\\
&& \nonumber
+ {\cal F}(b_1 x) \,{\cal F}(b_3 x) \,
\left[b_0 + b_2 {\cal F}(\varepsilon x)\right]\,,\\
\label{svm}
\bar\sigma_V(x) & = & \frac{1}{x+\bar m^2}\, \left[ 1 - {\cal F}(2 (x+\bar
m^2))\right]\,,
\end{eqnarray}
with ${\cal F}(y) = (1-{\rm e}^{-y})/y$, $x=p^2/\lambda^2$, $\bar m$ =
$m/\lambda$, $\bar\sigma_S(x) = \lambda\,\sigma_S(p^2)$ and $\bar\sigma_V(x)
= \lambda^2\,\sigma_V(p^2)$.  The mass-scale, $\lambda=0.566\,$GeV, and
dimensionless parameter values:\footnote{
$\varepsilon=10^{-4}$ in Eq.~(\ref{ssm}) acts only to decouple the large- and
intermediate-$p^2$ domains.  The study used Landau gauge because it is a
fixed point of the QCD renormalisation group and $Z_2\approx 1$, even
nonperturbatively\cite{mr97}.}
\begin{equation}
\label{tableA} 
\begin{array}{ccccc}
   \bar m& b_0 & b_1 & b_2 & b_3 \\\hline
   0.00897 & 0.131 & 2.90 & 0.603 & 0.185 
\end{array}\;,
\end{equation}
were fixed in a least-squares fit to light-meson observables\cite{mark}, and
the dimensionless $u$-current-quark mass corresponds to
\begin{equation}
m^{1\,{\rm GeV}} = 5.1\,{\rm MeV}.
\end{equation}
This algebraic parametrisation combines the effects of confinement and
dynamical chiral symmetry breaking with free-particle behaviour at large
spacelike $p^2$~\cite{revbasti} and, as illustrated explicitly in Ref.\
\cite{cdresi}, its qualitative features have recently been confirmed in
simulations of lattice-QCD\cite{latticequark}.

Since the quark described by Eqs.\ (\ref{qprop})-(\ref{svm}) is confined
there is no solution of the mass-shell equation: $s+M^2(s)=0$.  However, in
calculations of observables, the Euclidean constituent-quark mass, $M^E$,
defined as the solution of $s=M^2(s)$, provides a realistic estimate of the
quark's active quasi-particle mass.  In the present example, 
\begin{equation}
\label{Ecqm}
M^E = 0.33 \,{\rm GeV}\,.
\end{equation}

The large value of the ratio: $M^E/m^{1\,{\rm GeV}} \approx 70$ is one signal
of the effect that the dynamical chiral symmetry breaking mechanism has on
light-quark propagation characteristics.  Another is the value of the
chiral-limit vacuum quark condensate and one advantage of using the algebraic
parametrisation is that it yields the following simple
expression\cite{mrpion}:
\begin{eqnarray}
-\langle \bar q q \rangle^0_\zeta & = &
\ln\left(\zeta^2/\Lambda_{\rm QCD}^2\right)\,\lambda^3\,\frac{3}{4\pi^2}\,
\frac{b_0}{b_1\,b_3}\,,
\end{eqnarray}
which makes plain the relation between the condensate and the chiral-limit
coefficient of the $1/p^4$-term in $\sigma_S(p)$\cite{mr97}.

The general form of the $\pi$-meson Bethe-Salpeter amplitude is
\begin{eqnarray}
\nonumber
\Gamma_\pi(k;Q) & = &  \gamma_5 \left[ i E_\pi(k;Q) + 
\gamma\cdot Q \,F_\pi(k;Q) \rule{0mm}{5mm}\right. \\
\nonumber
& & +\left. \rule{0mm}{5mm} \gamma\cdot k \,k \cdot Q\, G_\pi(k;Q) 
+ \sigma_{\mu\nu}\,k_\mu Q_\nu \,H_\pi(k;Q) 
\right]\,,\\
\label{genpibsa}
\end{eqnarray}
where the behaviour of the invariant functions is constrained to a large
extent by the axial-vector Ward-Takahashi identity\cite{mrt98,mr97}.  That
identity and numerical studies\cite{mr97} support the parametrisation
\begin{equation}
E_\pi(k;Q) = \frac{1}{N_\pi}\,B_\pi(k^2)\,,
\end{equation}
where $B_\pi$ is obtained from Eqs.~(\ref{qprop})-(\ref{svm}), evaluated
using $\bar m=0$ and 
\begin{equation}
b_0\to b_0^\pi=0.19
\end{equation}
with the other parameters unchanged, and:
\begin{eqnarray}
\label{bsaF}
F_\pi(k;Q) & = & E_\pi(k;Q)/(110 f_\pi)\,; \\
\label{bsaG}
G_\pi(k;Q) & = & 2 F_\pi(k;Q)/[k^2+M_{\rm UV}^2]\,, 
\end{eqnarray}
$M_{\rm UV}=10\,\Lambda_{\rm QCD}$; and $H_\pi(k;Q)\equiv 0$.  The amplitude
is normalised canonically and consistent with the impulse approximation
$(k_\pm = k \pm Q/2)$:
\begin{eqnarray}
\nonumber \lefteqn{ Q_\mu = 
\int\!\frac{d^4 k}{(2\pi)^4}\left\{\rule{0mm}{5mm}
{\rm tr}_{\not F} \left[ 
\bar{\Gamma}_\pi(k;-Q) \frac{\partial S(k_+)}{\!\!\!\!\!\!\partial Q_\mu} 
{\Gamma}_\pi(k;Q) S(k_-) \right]
\right. 
} \\
& & \nonumber \left.  
\;\;\;\;\;\;\;\;\;\;\;\;\;\;\;\;\;\;\;
 + {\rm tr}_{\not F} \left[ 
\bar{\Gamma}_\pi(k;-Q) S(k_+) {\Gamma}_\pi(k;Q) 
        \frac{\partial S(k_-)}{\!\!\!\!\!\!\partial Q_\mu}\right]
\rule{0mm}{5mm}\right\}  \,,\\
\label{pinorm}
\end{eqnarray}
with the trace over flavour indices omitted, and this fixes $N_\pi$.  The
often-neglected pseudovector elements: $F_\pi$, $G_\pi$, are
crucial\cite{mrpion} to recovering the $1/q^2$ behaviour of the
electromagnetic pion form factor at large spacelike-$q^2$ exhibited in
perturbative QCD analyses.  However, models efficacious for low pion-energy
can be constructed without them\cite{pipi,mark,pichowsky2,mishasvy,cdrpion}.

In Ref.\ \cite{mrt98} a pseudoscalar meson mass formula was derived that, in
the limit of small current-quark masses, reproduces what is commonly known as
the Gell-Mann--Oakes--Renner relation, and also has an important corollary
applicable to mesons containing heavy-quarks\cite{mishasvy,hqcorollary}.  In
the case of the pion it gives
\begin{equation}
f_\pi^2\,m_\pi^2 = - 2 \,m^{\zeta} \, \langle \bar q q
\rangle^\pi_{\zeta}\,,
\end{equation}
with the parametrisations yielding an algebraic expression for the
``in-pion'' condensate:
\begin{equation}
- \langle \bar q q \rangle^\pi_{\zeta} =
\ln\left(\zeta^2/\Lambda_{\rm QCD}^2\right)\,\lambda^3\,\frac{3}{4\pi^2}\,
\frac{b_0^\pi}{b_1\,b_3}\,,
\end{equation}
and where the leptonic decay constant is obtained from
\begin{eqnarray}
\label{caint}
f_\pi Q_\mu = N_c\,{\rm tr}_D\int\frac{d^4 k}{(2\pi)^4}\, \gamma_5 \gamma_\mu
S(k_+) \Gamma_\pi(k;Q) S(k_-)\,.
\end{eqnarray}

It remains to specify the dressed-quark-photon vertex.  The manner in which
an Abelian gauge boson couples to a dressed-fermion has been much studied and
a range of qualitative constraints have been elucidated\cite{ayse97}.  This
research supports an {\it Ansatz}\cite{bc80}
\begin{eqnarray}
\nonumber
\lefteqn{i\Gamma_\mu(\ell_1,\ell_2) = 
i\Sigma_A(\ell_1^2,\ell_2^2)\,\gamma_\mu }\\ 
& & \nonumber 
+
(\ell_1+\ell_2)_\mu\,\left[\sfrac{1}{2}i\gamma\cdot (\ell_1+\ell_2) \,
\Delta_A(\ell_1^2,\ell_2^2) + \Delta_B(\ell_1^2,\ell_2^2)\right]\,;\\
&& \label{bcvtx}\\\
&&  \Sigma_F(\ell_1^2,\ell_2^2) = \sfrac{1}{2}\,[F(\ell_1^2)+F(\ell_2^2)]\,,\\
&& \Delta_F(\ell_1^2,\ell_2^2) =
\frac{F(\ell_1^2)-F(\ell_2^2)}{\ell_1^2-\ell_2^2}\,,
\end{eqnarray}
where $F= A, B$ are the scalar functions in Eq.\ (\ref{qprop}), which
preserves many of the constraints, important among them the vector
Ward-Takahashi identity, and is expressed solely in terms of the
dressed-quark propagator.  In concert with the algebraic parametrisations
described above, it has been widely used in the study of electromagnetic
processes and is phenomenologically efficacious; e.g., providing for the
parameter-independent realisation of ``anomalous'' photon-hadron
couplings\cite{cdrpion,ganomalies}.  For these reasons, we employ Eq.\
(\ref{bcvtx}) herein.  Nevertheless, significant progress has recently been
made\cite{mtpion,marisvtx,mariskaon} with the direct calculation of
$\Gamma_\mu$ in DSE models of QCD and those studies will provide the basis
for improved {\it Ans\"atze} in the future.

The elements described herein yield the following calculated values for a
selected, illustrative range of light-hadron observables:
\begin{eqnarray}
&& \nonumber
\begin{array}{c|ll}
        & {\rm Calc.} & {\rm Obs.} \\\hline
f_\pi ({\rm GeV})       & 0.090         &  0.092\cite{pdg00}    \\
m_\pi                   & 0.139         &  0.138\cite{pdg00}    \\
(-\langle \bar q q \rangle^0_\zeta)^{1/3} 
                        & 0.221         &  0.236\cite{derek}    \\
(-\langle \bar q q \rangle^\pi_{\zeta})^{1/3} 
                        & 0.250         &  0.245\cite{mr97}     \\
r_\pi ({\rm fm})        & 0.56          &  0.663\cite{amend}    \\
r_p\cite{martinnucleon}                     
                        &  0.78         &  0.87                 \\
-r_n^2 ({\rm fm}^2)\cite{martinnucleon}  &(0.33)^2      & (0.34)^2
\end{array}\\
\label{oldresults}
\end{eqnarray}
with $r_\pi$ calculated in impulse approximation (see, e.g., Refs.\
\cite{mrpion,piloop}) and the condensates evaluated at $\zeta=1\,$GeV using a
$3$-flavour value of $\Lambda_{\rm QCD}=0.242\,$GeV.

\section{Calculated Distribution Functions}
We can now proceed with the evaluation of the one-dimensional integral that
yields $u_v(x)$ via Eqs.\ (\ref{Wnew}) and (\ref{uvx}).  That calculation
requires a determination of the valence-quark mass, $\check M$.  The
parametrisation of the dressed-quark propagator is confining and does not
admit a solution of $\check M=M(-\check M^2)$.  However, in inclusive deep
inelastic scattering, confinement is recovered through incoherent
hadronisation after the dissolution of the bound state and we introduce that
aspect herein by adopting a quasiparticle representation: $\Delta(k^2)=
1/[k^2+\check M^2]$ in Eq.\ (\ref{SDN}), with $\check M$ determined by
requiring that Eq.\ (\ref{uvnorm}) is satisfied.  It is valence-quarks of
this mass that populate the pion.  This is an internally consistent
prescription if $\check M \approx M^E$.

\begin{figure}[t]
\centering{\
\epsfig{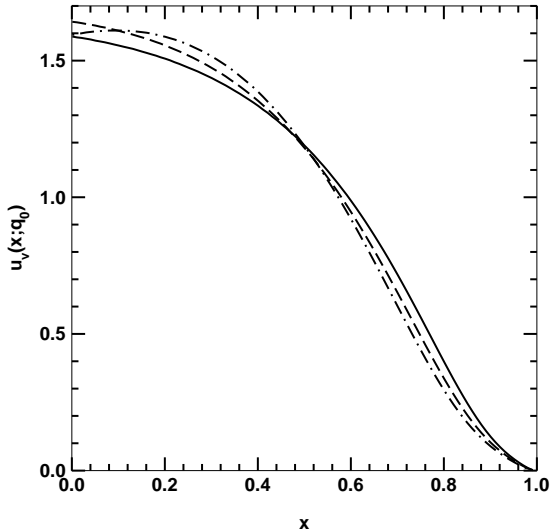}}\vspace*{1ex}
\caption{\label{uvxpic} Solid line: $u_v(x;q_0)$ calculated using the DSE
elements and parameters described in Sec.\ \protect\ref{DSEmodel}.  The
valence-quark mass is $\check M=0.30\,$GeV and the resolving scale
$q_0=0.54\,{\rm GeV}=1/(0.37\,{\rm fm})$ is fixed as described in connection
with Eq.\ (\protect\ref{q0scale}).  Dashed line: $u_v(x;q_0)$ calculated in
the absence of the pseudovector components of the pion's Bethe-Salpeter
amplitude; i.e., $F=0=G$ in Eq.\ (\protect\ref{genpibsa}) instead of Eqs.\
(\protect\ref{bsaF}) and (\protect\ref{bsaG}).  Dot-dashed line: distribution
calculated with $m_\pi\to 0.1\,m_\pi$, $\check M=0.36\,$GeV.}
\end{figure}

Our calculated form of $u_v(x;q_0)$ is depicted in Fig.\ \ref{uvxpic}.  It
vanishes at $x=1$, in accordance with the kinematic constraint expressed in
Eq.\ (\ref{mumin}), and corresponds to a finite value of $F_1(x=0)$, which is
a signal of the absence of sea-quark contamination.  Furthermore
$u_v(x=0;q_0)\neq 0$, which is as it should be since there is no constraint
that requires it to vanish at this point.  Unsurprisingly for a light bound
state of heavy constituents, the shape of the distribution is characteristic
of a strongly bound system\cite{piller}: cf.\ for a weakly bound system
$u_v(x) \approx \delta(x-\case{1}{2})$.  

The area under each of the curves in Fig.\ \ref{uvxpic} is one and requiring
that for $u_v(x;q_0)$ yields a calculated valence-quark mass of
\begin{equation}
\check M = 0.30\,{\rm GeV}\,,
\end{equation}
which is within 10\% of this model's Euclidean con\-sti\-tu\-ent-quark mass,
Eq.\ (\ref{Ecqm}).  The value of $\check M$ affects the position of the peak
in $x u_v(x)$: increasing $\check M$ shifts the peak to lower $x$.

The average momentum-fraction carried by the valence-quarks at this resolving
scale is
\begin{equation}
\int_0^1\!dx\,x\,[u_v(x;q_0)+\bar d_v(x;q_0)] = 0.71\,,
\end{equation}
with the remainder carried by the gluons that effect the binding of the pion
bound state, which are invisible to the electromagnetic probe.  The second
and third moments of the distribution are
\begin{eqnarray}
\langle x^2\rangle_{q_0}  = 0.18\,, & \;& 
\langle x^3\rangle_{q_0}  = 0.10\,.
\end{eqnarray}

To determine the resolving scale, $q_0$, we employ the $3$-flavour,
leading-order, nonsinglet renormalisation group (evolution) equations (see,
e.g., Ref.\ \cite{pdg00}) to evolve the distribution in Fig.\ \ref{uvxpic} up
to $q=2\,$GeV, and require agreement between the first and second moments of
our evolved distribution and those calculated from the phenomenological fits
of Ref.\ \cite{sutton}.  With
\begin{equation}
\label{q0scale}
q_0 = 0.54\,{\rm GeV} = 1/(0.37\,{\rm fm})
\end{equation}
\begin{figure}[t]
\centering{\
\epsfig{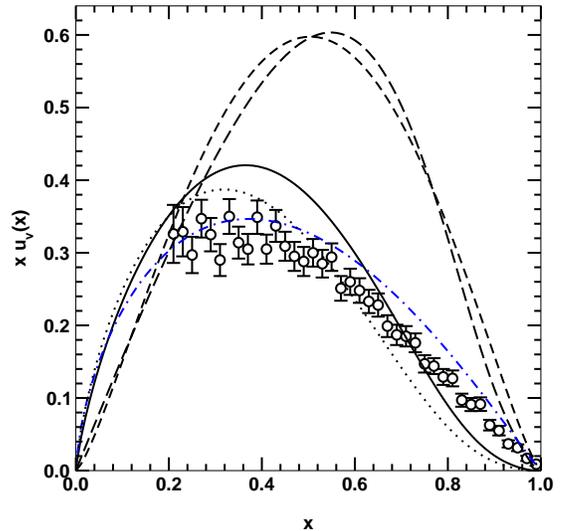}}\vspace*{1ex}
\caption{\label{xuvxpic} Dashed line: $x u_v(x;q_0)$; Short-dashed line: fit
of Eqs.\ (\protect\ref{fitxuvx}) and (\protect\ref{fitparams}); Solid line:
the evolved distribution, $x u_v(x;q=2\,{\rm GeV})$; Dotted line: $x
u_v(x;q=4.05\,{\rm GeV})$, evolved with a $4$-flavour value of $\Lambda_{\rm
QCD}= 0.204\,$GeV; and Dot-dashed line: the phenomenological fit of Ref.\
\protect\cite{sutton}.  The data are from Ref.\ \protect\cite{DYexp2},
obtained with an invariant $\mu^+ \mu^-$-mass $>$ $4.05\,$GeV and inferred
from the differential pion-nucleon Drell-Yan cross section using simple
distribution parametrisations of the type in Eq.\ (\protect\ref{fitxuvx}).
This data was part of the set employed in the fit of Ref.\
\protect\cite{sutton}.  The fits of Eqs.\ (\protect\ref{fit94}) and
(\protect\ref{fitparamsnew}) are not depicted: they are almost
indistinguishable from our calculated curves.}
\end{figure}
\hspace*{-\parindent}we obtain at $q=2\,$GeV
\begin{equation}
\begin{array}{l|lll}
                &       \langle x\rangle_q   & \langle x^2 \rangle_q & 
        \langle x^3 \rangle_q \\\hline
{\rm Calc.}     &        0.24                   & 0.098 & 0.049      \\
{\rm Fit}\protect\cite{sutton}& 0.24\pm 0.01    & 0.10\pm 0.01 &  
                0.058 \pm 0.004      \\
{\rm Latt.}\protect\cite{lattice}
                & 0.27 \pm 0.01                 & 0.11 \pm 0.3 & 
                0.048 \pm 0.020 
\end{array}
\end{equation}
at which point the valence-quarks carry a momentum-fraction of $0.49$.  At
$q_0$, $\alpha_s/(2\pi) = 0.13$: this is the combination that appears in the
evolution equation.  (NB.\ At $0.5\,q_0$, $\alpha_s/(2\pi) = 0.68$.)

The original and evolved distributions are depicted in Fig.\ \ref{xuvxpic}.
The evident accentuation via evolution of the convex-up behaviour of the
distribution near $x= 1$ is characteristic of the renormalisation group
equations, which populate the sea-quark distribution at small-$x$ at the
expense of large-$x$ valence-quarks.

A fit to $x \,u_v(x;q)$, adequate for the rapid estimation of moments, is
provided by the simple data fitting form employed in Refs.\
\cite{DYexp2,sutton}:
\begin{equation}
\label{fitxuvx}
x\,u_v^{\rm mom}(x;q) = x^\alpha \, (1-x)^\beta\,
\frac{\Gamma(1+\alpha+\beta)}{\Gamma(\alpha)\,\Gamma(1+\beta)}\,,
\end{equation}
with in our case
\begin{equation}
\begin{array}{c|lll}
q \,({\rm GeV}) & 0.57 & 2.0   & 4.05 \\\hline
\alpha                  & 1.34 & 0.92 &  0.84    \\
\beta                   & 1.31 & 1.80  & 1.98  \;.    
\end{array}\\
\label{fitparams}
\end{equation}
The moments of $u_v^{\rm mom}(x;q)$ are given by
\begin{equation}
\langle x^n \rangle_q = \prod_{i=1}^n\,\frac{ i+ \alpha
-1}{i+\alpha+\beta}\,.
\end{equation}
We emphasise, as is evident in Fig.\ \ref{xuvxpic}, that Eq.\ (\ref{fitxuvx})
is not a good pointwise approximation to our calculated form of $x \,
u_v(x;q_0)$ and, furthermore, Eq.\ (\ref{fitxuvx}) divided-by $x$ provides a
very poor pointwise approximation to $u_v(x;q_0)$ for $x<0.5$.  

An alternative parametrisation has also been employed\cite{mrs94} in fitting
data:
\begin{equation}
\label{fit94}
x\, u_v^{\rm cu}(x) =
A_u\,x^{\eta_1}\,(1-x)^{\eta_2}\,(1-\epsilon_u\,\sqrt{x}+\gamma_u \,x)\,,
\end{equation}
with $A_u$ fixed by Eq.\ (\ref{uvnorm}).  This, with the parameter values:
\begin{equation}
\begin{array}{c|rrl}
q \,({\rm GeV}) & 0.57 & 2.0   & 4.05 \\\hline
A_u             & 11.24 & 4.25  & 3.56    \\
\eta_1          &  1.43 & 0.97  & 0.89    \\
\eta_2          &  1.90 & 2.43  & 2.61     \\
\epsilon_u      &  2.44 & 1.82  & 1.62    \\
\gamma_u        &  2.54 & 2.46  & 2.30\;,
\end{array}\\
\label{fitparamsnew}
\end{equation}
provides a pointwise accurate interpolation of the calculated forms of $x
\,u_v(x)$ depicted in Fig.\ \ref{xuvxpic}: on the scale of this figure, the
fit and our calculation are barely distinguishable.  Furthermore, for $x\geq
0.2$, $u_v^{\rm cu}(x;q_0)$ obtained from Eq.\ (\ref{fit94}) even provides a
good pointwise approximation to $u_v(x;q_0)$ in Fig.\ \ref{uvxpic}.  The
difference between Eqs.\ (\ref{fitxuvx}) and (\ref{fit94}) is the improved
capacity of the latter to accommodate convexity in the parton distributions,
which is a signal feature of our calculation.

Particular regularisations of the Nambu--Jona-Lasinio model
yield\protect\cite{arriola,bentz} a distribution with the functional form
$u_v^\theta(x;q_0^{\rm NJL})=\theta(x)\,\theta(1-x)$, which corresponds to
valence-quarks carrying each and every fraction of the pion's momentum with
equal probability.  That result is an artefact, arising from the
representation of the pion bound state by a momentum-{\it independent}
Bethe-Salpeter amplitude; i.e., from representing the pion as a
point-particle, which is a necessary consequence of the model's
momentum-independent interaction.  In this case, beginning at $q_0^{\rm NJL}
= 0.35$, for which $\alpha/(2\pi) = 0.31$, $x\, u_v^\theta(x;q_0^{\rm NJL})$
evolved to $q=2\,$GeV is pointwise very well described by Eq.\
(\ref{fitxuvx}) with $\alpha=0.67$, $\beta=1.13$, cf.\ the values in Ref.\
\cite{sutton}: $\alpha= 0.64\pm 0.03$, $\beta = 1.15\pm 0.02$.

We infer from this result and the discussion above that the fitting form in
Eq.\ (\ref{fitxuvx}) is inadequate: it is continuously connected to a
distribution that lacks dynamical content and is unable to represent that
structure in the distributions which characterises dynamics.  These
observations yield insight into the efficacy of the updated fitting
form\cite{mrs94} in Eq.\ (\ref{fit94}).

\section{Epilogue}
We calculated the valence-quark distribution in the pion using a
Dyson-Schwinger equation (DSE) model that provides a good description of a
wide range of hadron observables: Figs.\ \ref{uvxpic} and \ref{xuvxpic}
summarise our new results.  The DSE model describes valence-quarks with an
active mass $\check M = 0.30\,$GeV at a resolving scale
$q_0=0.54\,$GeV$\,=1/(0.37\,{\rm fm})$, and with this value of $q_0$ the
evolution to values of the momentum scale relevant to contemporary
experiments yields a distribution whose low moments agree with the values
obtained in lattice simulations and from a phenomenological fit.

There is a pointwise difference between our calculated distribution and the
form used hitherto in parametrising pion data, as evident in Fig.\
\ref{xuvxpic}.  That discrepancy can also be observed in related covariant
calculations\cite{toki,bentz,dorokhov}.  From the information currently
available, its origin does not appear to lie in model details, and we note
that any assumed fitting form with little or no convexity in the vicinity of
$x=1$, which is adapted to a body of data at a given scale, $q_0^2$, will
necessarily become convex-up under evolution to an higher scale, $q^2>q_0^2$.
Part or all of this discrepancy may therefore be attributable to the
restricted function space used thus far in parametrising pion data.  That
possibility is supported by the capacity of an updated fitting form, used for
the proton, to accommodate the structure evident in our calculations.
However, we cannot rule out the possibility that the discrepancy may point to
an as yet overlooked shortcoming in the application of models to the
calculation of distribution functions.

Our calculation is a prototype.  It can be improved; e.g., by using direct
numerical solutions of the quark-DSE and meson Bethe-Salpeter equations, as
in recent calculations of the light-meson electromagnetic form
factors\cite{mariskaon}, and/or incorporating sea-quark contributions at the
soft-scale, $q_0$.  It can also be extended to more robust ``targets,'' such
as the nucleon, using models like those of Refs.\
\cite{martinnucleon,reinhardnucleon}.

\section*{Acknowledgments}
We acknowedge constructive conversations with J.C.R.~Bloch, A.~Drago and
R.J.~Holt.  C.D.R. is grateful for the support and hospitality of the Erwin
Schr\"odinger Institute for Mathematical Physics, Vienna, where part of this
work was completed, and S.M.S. acknowledges financial support from the
A.v.~Humboldt foundation.  This work was supported by the US Department of
Energy, Nuclear Physics Division, under contract no. W-31-109-ENG-38, the
National Science Foundation, under grant no. INT-9603385, and benefited from
the resources of the National Energy Research Scientific Computing Center.



\end{document}